\documentclass[aps, prb, twocolumn, amssymb, amsmath, showpacs, superscriptaddress]{revtex4-1}

\usepackage{bm}
\usepackage{times}
\usepackage{graphicx}
\usepackage[colorlinks,citecolor=blue,linkcolor=blue]{hyperref}
\usepackage{float}

\begin{document}

\title{Effect of modulations of doping and strain on the electron transport in monolayer ${\bf{MoS}_2}$}

\author{Yanfeng Ge$^\ddagger$}
\affiliation{School of Physics, Beijing Institute of Technology, Beijing 100081, China}
\author{Wenhui Wan$^\ddagger$}
\affiliation{School of Physics, Beijing Institute of Technology, Beijing 100081, China}
\author{Wanxiang Feng}\email{wxfeng@bit.edu.cn}
\affiliation{School of Physics, Beijing Institute of Technology, Beijing 100081, China}
\author{Di Xiao}
\affiliation{Department of Physics, Carnegie Mellon University, Pittsburgh, Pennsylvania 15213, USA}
\author{Yugui Yao}\email{ygyao@bit.edu.cn}
\thanks{\\$^\ddagger$These authors contributed equally to this work}
\affiliation{School of Physics, Beijing Institute of Technology, Beijing 100081, China}

\date{\today}

\begin{abstract}
The doping and strain effects on the electron transport of monolayer $\rm{MoS}_{2}$ are systematically investigated using the first-principles calculations with Boltzmann transport theory. We estimate the mobility has a maximum 275 cm$^2$/(V$\cdot$s) in the low doping level under the strain-free condition. The applying a small strain ($\sim$3\%) can improve the maximum mobility to 1150 cm$^2$/(V$\cdot$s) and the strain effect is more significant in the high doping level. We demonstrate that the electric resistance mainly due to the electron transition between K and Q valleys scattered by the M momentum phonons. However, the strain can effectively suppress this type of electron-phonon coupling by changing the energy difference between the K and Q valleys. This sensitivity of mobility to the external strain may direct the improving electron transport of $\rm{MoS}_{2}$.
\end{abstract}


\maketitle

\section{INTRODUCTION}
After the initial boom in graphene research, recent years have seen a surge of interest in other two-dimensional (2D) atomic crystals~\cite{geim2013}. Among them, molybdenum disulfide ($\rm{MoS}_{2}$), a prototypical transition metal dichalcogenide, has attracted great attention due to its excellent electronic and optical properties~\cite{wang2012,ferdows2013,Feng2012,Conley2013,Gomez2013}. Similar to graphite, bulk $\rm{MoS}_{2}$ consists of vertically stacked layers that are loosely coupled via the Van der Waals interaction.  When shaped into monolayer, the energy gap changes from indirect to direct ($\sim$1.9eV), right in the visible frequency range~\cite{Splendiani2010,mak2010}, allowing applications such as transistors~\cite{radisavljevic2011}, photodetectors and electroluminescent devices. In addition, due to the strong spin-orbit interaction and broken inversion symmetry, monolayer $\rm{MoS}_{2}$ also exhibit novel valley and spin physics as demonstrated recently~\cite{xiao2012,zeng2012,mak2012,cao2012}.

To realize its application potential in multi-functional electronic devices, it is essential to understand the transport mechanisms in $\rm{MoS}_{2}$.  Compared with graphene~\cite{geim2007,chen2008}, the mobility of pristine $\rm{MoS}_{2}$ is rather low, typically on the order of 10 cm$^2$/(V$\cdot$s) at room temperature.  Higher mobilities can be achieved by gate dielectric engineering to effectively screen the Coulomb scattering on charged impurities and suppress electron-phonon scattering, values in the range from 200 to 700 cm$^2$/(V$\cdot$s) have been reported~\cite{radisavljevic2011,Wang2012a,bao2013,das2013}. On the other hand, theoretical calculations of the phonon-limited mobility, using the deformation potential approximation or full band Monte Carlo simulation,  have placed an intrinsic limit in the range of $130\sim410$ cm$^2$/(V$\cdot$s) at room temperature~\cite{Kaasbjerg2012,li2013}, which are close to experimental values. This suggests that further improvement of the mobility must come from better control of electron-phonon coupling.

Strain engineering~\cite{Feng2012,Conley2013,Gomez2013} and electron-doping~\cite{ye2012,Taniguchi2012,Radisavljevic2013,Perera2013} have been successfully used to improve the performance of $\rm{MoS}_{2}$, such as energy gap and superconductivity.
Here, we have systematically investigated the electron-doping and strain dependence of the transport electron-phonon coupling constant $\lambda_{tr}$ and the phonon-limited mobility $\mu$ in monolayer $\rm{MoS}_{2}$ based on first-principles calculations and the Boltzmann transport theory.
It is found that $\lambda_{tr}$ increases with increasing doping concentration $n_\text{2D}$, reaches the maximum at $n_\text{2D} = 1.7 \times 10^{14}$ cm$^{-2}$ before starts decreasing.  Remarkably, $\lambda_{tr}$ can be significantly reduced by applying a small amount of strain.  This is due to the change of energy difference between the K and Q valleys, which effectively suppresses the inter-valley scattering~\cite{li2013,song2013}.  We show that even with 3\% strain the mobility can be increased from 275 to 1150 cm$^2$/(V$\cdot$s). Our results provide an effective method to modulate the electron transport of $\rm{MoS}_{2}$
by external strain.

\section{Methodology}

Our calculation is based on the semiclassical Boltzmann transport theory. Details of theory can be found in Ref.~\onlinecite{allen1978}. The key quantity is the transport electron-phonon coupling strength $\lambda^{\mathbf{q},\nu}_{tr}$for wavevector $\mathbf{q}$ and mode index $\nu$, defined by~\cite{allen1978}
\begin{eqnarray}
\begin{aligned}
\lambda^{\mathbf{q},\nu}_{tr}=\frac{2}{{N_F}{N_k}\omega_{\mathbf{q},\nu}}{\sum_{\substack{\mathbf{k},\rm i, \rm j}}}
{\vert{M^{\nu}_{\mathbf{k} \rm i,(\mathbf{k}+\mathbf{q}) \rm j}}\vert}^2
\delta(\epsilon_{\mathbf{k}\rm i}-\epsilon_{F})\\
\delta(\epsilon_{(\mathbf{k}+\mathbf{q})\rm j}-\epsilon_{F})\eta_{\mathbf{k}\rm i,(\mathbf{k}+\mathbf{q}) \rm j},
\end{aligned}
\label{eq:epcq}
\end{eqnarray}
where $\rm{M}^{\nu}_{\mathbf{k}\rm i,(\mathbf{k}+\mathbf{q})\rm j}$ is the electron-phonon interaction matrix element,
\begin{eqnarray}
\rm{M}^{\nu}_{\mathbf{k}\rm i,(\mathbf{k}+\mathbf{q})\rm j}=\sqrt{\frac{\hbar}{2M \omega_{\mathbf{q},\nu}}} \langle\mathbf k i| \delta^{\mathbf{q},\nu}V_{SCF}|(\mathbf k+\mathbf q)j \rangle,
\label{eq:m}
\end{eqnarray}
and the efficiency factor,
\begin{eqnarray}
\eta_{\mathbf{k}\rm i,(\mathbf{k}+\mathbf{q})\rm j}=
1-\frac{{\mathbf v_{\mathbf k \rm i}}\cdot{\mathbf v_{(\mathbf k+\mathbf q) \rm j}}}{{\vert{\mathbf v_{\mathbf k \rm i}}\vert}^2}.
\label{eq:eta}
\end{eqnarray}
In Eq.~(\ref{eq:epcq}), $N_F$ is the density of state at the Fermi surface, $N_k$ is the total number of $\mathbf k$ points, $\epsilon_{\mathbf{k}\rm i}$ and $\mathbf v_{\mathbf k \rm i}$ are the band energy and group velocity of the Bloch electrons, respectively. In Eq.~(\ref{eq:m}), M is the atomic mass, and $\delta^{\mathbf{q},\nu} \rm V_{ \rm SCF}$ is the derivative of the self-consistent effective potential with respect to atomic displacement associated with the phonon from branch $\nu$ with the wave vector $\mathbf{q}$ and frequency $\omega_{\mathbf{q},\nu}$. The efficiency factor $\eta_{\mathbf{k}\rm i,(\mathbf{k}+\mathbf{q})\rm j}$ shows that only backward scattering will contribute to the resistance.

Similar to the relation between the Eliashberg function and the superconducting electron-phonon coupling constant, the transport spectral function $\alpha^2_{tr}\rm{F}(\omega)$and the transport electron-phonon coupling constant $\lambda_{tr}$ can be obtained by~\cite{allen1978},
\begin{eqnarray}
\alpha^2_{tr}\rm{F}(\omega)&=&\frac{1}{2N_q}{\sum_{\substack{\mathbf{q},\nu}}}
\lambda^{\mathbf{q},\nu}_{tr}\omega_{\mathbf{q},\nu}\delta(\omega-\omega_{\mathbf{q},\nu}),
\label{eq:a2fw}
\end{eqnarray}
\begin{eqnarray}
\lambda_{tr}=2\int^{\infty}_{0}\omega^{-1}\alpha^2_{tr}\rm{F}(\omega)\rm{d}\omega,
\label{eq:lambda}
\end{eqnarray}
where $N_q$ is the total number of $\mathbf q$ points.

Finally, using transport spectral function $\alpha^2_{tr}\rm{F}(\omega)$, the relaxation time $\tau$ can be derived by solving the Boltzmann equation in the lowest-order variational approximation (LOVA) as,
\begin{eqnarray}
\begin{aligned}
\tau^{-1}=
(\frac{\textstyle4\pi{k_B}T}{\hbar})
\int\frac{d\omega}{\omega}\frac{\tilde{\omega}^2}{\sinh^2\tilde{\omega}}\alpha^2_{tr}\rm{F}(\omega),
\end{aligned}
\label{eq:tau}
\end{eqnarray}
\begin{eqnarray}
\begin{aligned}
\frac{\tilde{\omega}^2}{\sinh^2 \tilde{\omega }}=\frac{\omega }{2k_B^2T^2}\int_{-\infty }^{\infty }d\epsilon \int_{-\infty }^{\infty }{d\epsilon'f(\epsilon)}[1-f(\epsilon')]\\
\{[N(\omega )+1]\delta (\epsilon -\epsilon'-\hbar \omega )
+N(\omega )\delta (\epsilon -\epsilon'+\hbar \omega )\}
\label{eq:sinhw}
\end{aligned}
\end{eqnarray}
where $\tilde{\omega}=\hbar\omega/(2{k_B}T)$. $f(\epsilon)$ and $N(\omega)$ are the Fermi-Dirac and the Bose-Einstein distribution function, respectively.
Moreover, the influence of impurities and electron-electron scattering are not included evidently in our calculations, which are subject to further investigations.

Utilizing the transport electron-phonon coupling, the temperature dependence of mobility ${\mu}(T)$ can be obtained by
\begin{eqnarray}
\mu(T)=\frac{2e {\emph N_ \emph F} \langle{v^2_x}\rangle }{n_{\rm 2D}\textstyle{S_{cell}}} \tau,
\end{eqnarray}
where $\langle{v^2_x}\rangle$ is the average square of the Fermi velocity along the $x$ direction, $S_{cell}$ is the area of unit cell.
\begin{figure}[htbp!]
\includegraphics[width=0.5\textwidth,height=4.3in]{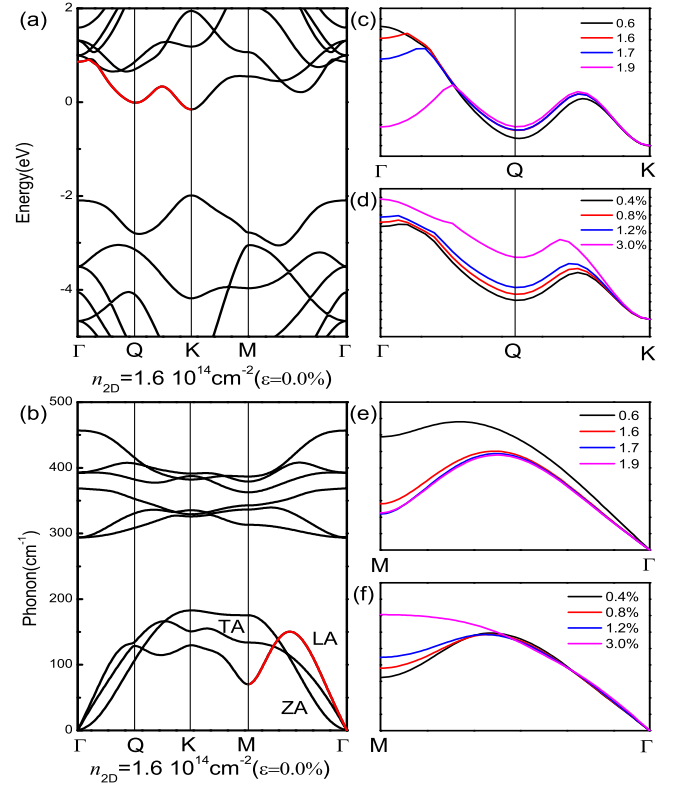}
\caption{(Color online) The band structure (a) and phonon dispersion (b) of strain-free monolayer $\rm{MoS}_2$ at doping concentration of $n_{\rm{2D}} = 1.6 \times 10^{14}$ cm$^{-2}$. The doping effect ($n_{\rm{2D}} =0.6, 1.6, 1.7, \textrm{and } 1.9 \times 10^{14}$ cm$^{-2}$) at $\varepsilon=0\%$ (c) and the strain effect ($\varepsilon=0.4\%, 0.8\%, 1.2\%, \textrm{and } 3.0\%$) at $n_{\rm{2D}} = 1.6 \times 10^{14}$ cm$^{-2}$ (d) on the lowest conduction band along the $\Gamma$-$K$ direction (labeled by red color in (a)). The band structure at $\Gamma$ and Q points are obviously modulated by doping and strain, respectively. The doping effect (e) and the strain effect (f) on the LA phonon mode along the $M$-$\Gamma$ direction (labeled by red color in (b)).\label{fig:1}}
\end{figure}

Technical details of the calculations are as follows. All calculations in this work were carried out in the framework of density functional theory (DFT) with local-density approximation (LDA)~\cite{Perdew1981}, as implemented in the QUANTUM ESPRESSO package~\cite{Paolo2009}. The ion and electron interactions are treated with the norm-conserving pseudopotentials~\cite{Troullier1991}. The kinetic energy cutoff of 30 Ry and Monkhorst-Pack $k$-mesh of $32\times32\times1$ were used in all calculations of electronic properties. The atomic positions were relaxed fully with the energy convergence criteria of $10^{-5}$ Ry and the force convergence criteria of $10^{-4}$ Ry/a.u. In our slab model, a vacuum layer with 15 \AA \ was set to avoid the interactions between the adjacent atomic layers. The equilibrium lattice constant of the monolayer $\rm{MoS}_{2}$ was found to be $a_0 = 3.11$ \AA. The electron doping was achieved by increasing the valence charge and at the same time introducing the same amount of uniform background charge. The strain was introduced by adjusted the lattice constant $a$ of the monolayer $\rm{MoS}_{2}$ with the strain capacity $\varepsilon=(a-a_0)/a_0\times100\%$.
The phonon dispersion and electron-phonon coupling were calculated on a $16\times16\times1$ $q$-grid using the density functional perturbation theory (DFPT)~\cite{Baroni2001}. We have carefully checked the convergence of q-point sampling with the denser grid of $20\times20\times1$, while the difference is only below 1\%.

\section{Result and Discussion}
\subsection{Electronic structure and phonon dispersion}

\begin{figure*}[htbp!]
\centering
\includegraphics[width=\textwidth]{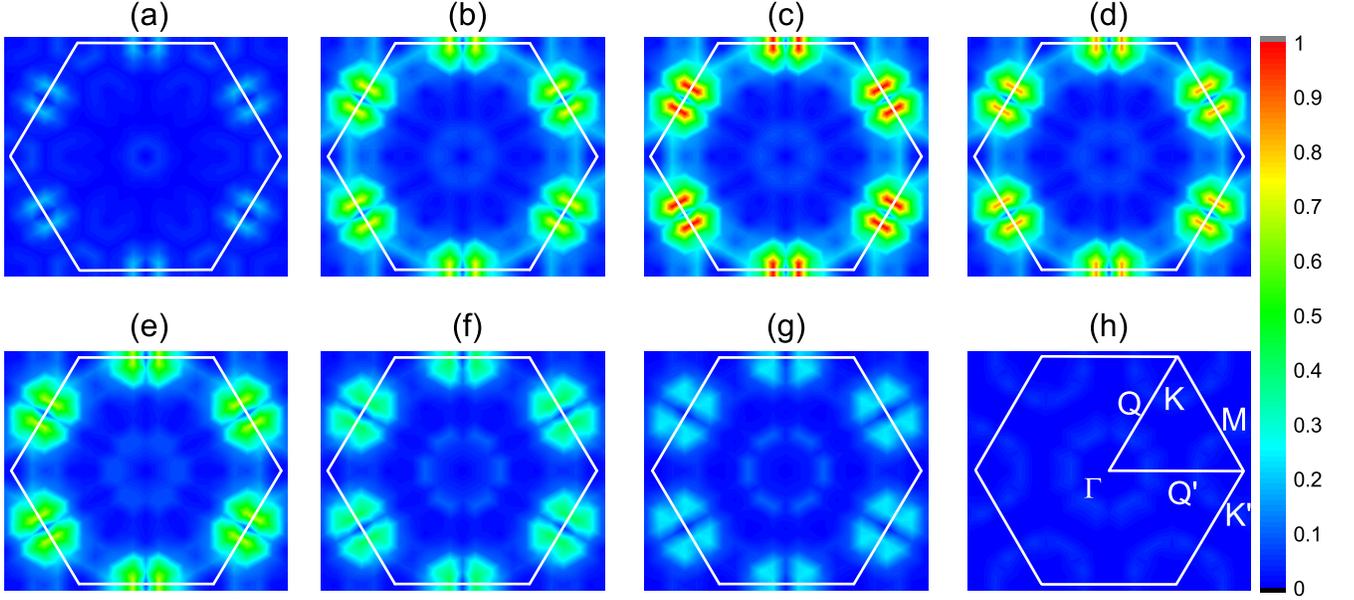}
\caption{(Color online) The map distributions of wave-vector-resolved electron-phonon coupling parameter $\lambda^{\mathbf{q}}_{tr}$ (a-d) at different doping concentrations $n_{\rm{2D}} = 0.6(\rm{a})$, $1.6(\rm{b})$, $1.7(\rm{c})$ \rm{and } $1.9(\rm{d}) \times 10^{14}$ cm$^{-2}$ with strain-free condition, respectively. The map distributions of $\lambda^{\mathbf{q}}_{tr}$ (e-h) at different tensile strains $\varepsilon = 0.4\%(\rm{e})$, $0.8\%(\rm{f})$, $1.2\%(\rm{g})$, \rm{and } $3.0\%(\rm{h})$ at the doping concentrations of $n_{\rm{2D}} = 1.7 \times 10^{14}$ cm$^{-2}$, respectively. \label{fig:2}}
\end{figure*}
Figure~\ref{fig:1} shows the electronic band structure and phonon dispersion of monolayer $\rm{MoS}_{2}$ with different electron doping concentrations and strains.  Central to our discussion is the existence of a second energy minima in the conduction band, called Q valley, which is about 100 meV higher than the conduction band minimum at the K valley, and approximately located at the halfway point of the $\Gamma$-K line. Because of the close proximity in energy, the inter-valley scattering between the K and Q valleys has a significant effect on transport~\cite{li2013,song2013}.  We note that while the conduction band at the $\Gamma$ point displays a strong dependence on the doping concentration, with a large drop after $n_{\rm{2D}} > 1.7 \times 10^{14}$ cm$^{-2}$, the band at the Q point shows little change with various concentrations [Fig.~\ref{fig:1}(c)]. On the other hand, the energy difference between the K and Q valleys (E$_{\text{KQ}}$) can be modified significantly by the strain [Fig.~\ref{fig:1}(d)], consistent with previous calculations~\cite{Chang2013}. The change of phonon dispersion also shows the important roles of the electron doping and strain.  The acoustic phonon branch (\textit{i.e.,} longitudinal acoustic mode) around the M point (denoted by $\rm{M}_p$ for simplicity) appears phonon softening with increasing doping concentration up to ${1.7} \times 10^{14}$ cm$^{-2}$.  However, for $n_{\rm{2D}} > 1.7 \times 10^{14}$ cm$^{-2}$ the phonon dispersion does not show any obvious change as $n_\text{2D}$ is varied [Fig.~\ref{fig:1}(e)].  The $\rm{M}_p$ phonons softening is also significantly suppressed by applying the strain [Fig.~\ref{fig:1}(f)].

\subsection{Electron-phonon coupling}
The strong dependence of both the electronic band structure and phonon dispersion on the doping and strain indicates that the electron-phonon coupling should show similar trend. To quantify this statement, we have calculated the transport spectral function $\alpha^2_{tr}\rm{F}(\omega)$ and transport electron-phonon coupling constant $\lambda_{tr}$.

The map distribution of wave-vector-resolved transport electron-phonon coupling parameter $\lambda^{\mathbf{q}}_{tr} = {\sum\limits_{\nu}}{\lambda^{\mathbf{q},\nu}_{tr}}$ as a function of doping concentrations in the absence of strain is shown in Figs.~\ref{fig:2}(a)-\ref{fig:2}(d).
One can see that the transport electron-phonon coupling around the M point is always the strongest, while those of K and $\Gamma$ points are rather small.
It indicates that $\rm{M}_p$ phonons is the dominating contributor to the transport electron-phonon coupling. The electronic transitions between the K and Q valleys scattered by $\text{M}_{\rm p}$ phonons~\cite{li2013} is described as $\text{K}_{\rm e}+\text{M}_{\rm p}\longleftrightarrow\text{Q}_{\rm e}$, where $\text{K}_{\rm e}$ ($\text{Q}_{\rm e}$) is the electronic state with the momentum K (Q). With the increase of doping concentration, the electrons occupying on the K and Q valleys increase and the frequency of phonons around the M point decreases [Fig.~\ref{fig:1}(e)]. Consequently, the coupling to $\text{M}_{\rm p}$ phonons increases sharply [Figs.~\ref{fig:2}(a)-\ref{fig:2}(c)]. However, it starts decreasing when $n_{\rm{2D}} > 1.7 \times 10^{14}$ cm$^{-2}$ [Fig.~\ref{fig:2}(d)]. One reason is the little dependence of the $\text{M}_{\rm p}$ phonons on $n_\text{2D}$ after it reaches $1.7 \times 10^{14}$ cm$^{-2}$ [Fig.~\ref{fig:1}(e)]. The other is the reduction of electrons in the Q valley, which arises from the lowering of the conduction band at the $\Gamma$ point [Fig.~\ref{fig:1}(c)].
Furthermore, the lowering of conduction band at $\Gamma$ point also results in the appearance of two more channels of electronic transitions, $\Gamma_{\rm e}+\text{Q}_{\rm p}\longleftrightarrow\text{Q}_{\rm e}$ and $\Gamma_{\rm e}+\text{K}_{\rm p}\longleftrightarrow\text{K}_{\rm e}$, which can be derived from the band structure and nesting function (see Appendix).
However, since the phonon frequencies at both K and Q points are higher than that of M point [Fig.~\ref{fig:1}(b)], according to the Eq.~(\ref{eq:epcq}), these two additional channels have much weaker electron-phonon coupling comparing with the $\text{K}_{\rm e}+\text{M}_{\rm p}\longleftrightarrow\text{Q}_{\rm e}$ channel [Fig.~\ref{fig:2}(d)].  More interestingly, $\lambda^{\mathbf{q}}_{tr}$ can be greatly suppressed once a small amount of strain is introduced [Figs.~\ref{fig:2}(e)-\ref{fig:2}(h)].  This is because of the hardening of the $\rm{M}_p$ phonons under strain condition and the increase of E$_{\text{KQ}}$ [Fig.~\ref{fig:1}(d)].
\begin{figure}[htbp!]
\includegraphics[width=0.5\textwidth]{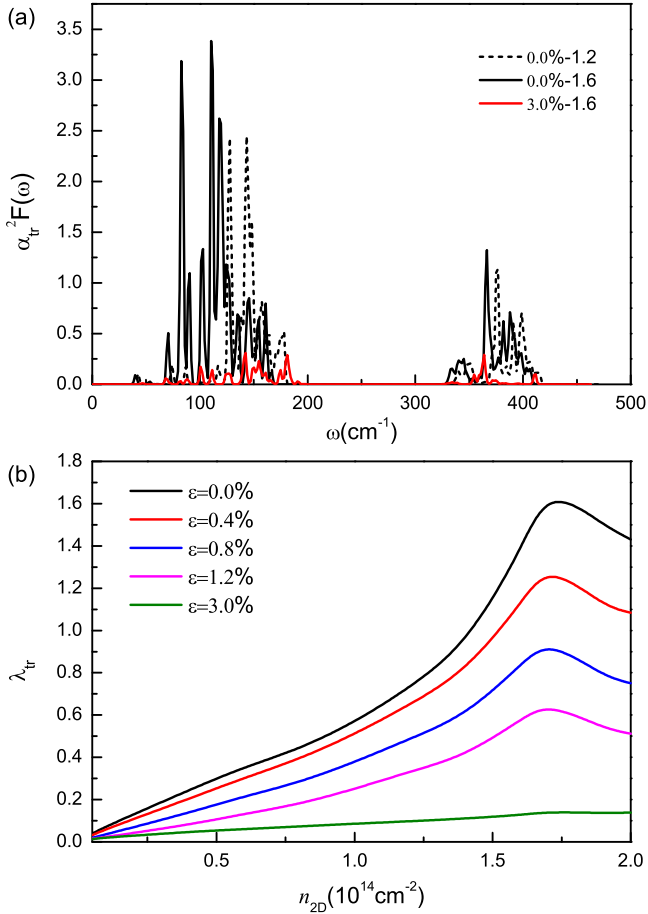}
\caption{(Color online) (a) Transport spectral function $\alpha^2_{tr}\rm{F}(\omega)$ at the doping concentration of $n_{\rm{2D}} = 1.2 \times 10^{14}$ cm$^{-2}$ with $\varepsilon = 0.0\% $ (black dash line) and $n_{\rm{2D}} = 1.6 \times 10^{14}$ cm$^{-2}$ with $\varepsilon = 0.0\%$ (black solid line) and $\varepsilon = 3.0\%$ (red solid line). (b) transport electron-phonon coupling constant $\lambda_{tr}$ as a function of the doping concentration $n_{\rm{2D}}$ at various strains of $\varepsilon$ = 0\%, 0.4\%, 0.8\%, 1.2\%, and 3.0\%. \label{fig:3}}
\end{figure}
\begin{figure}[htbp!]
\includegraphics[width=0.5\textwidth]{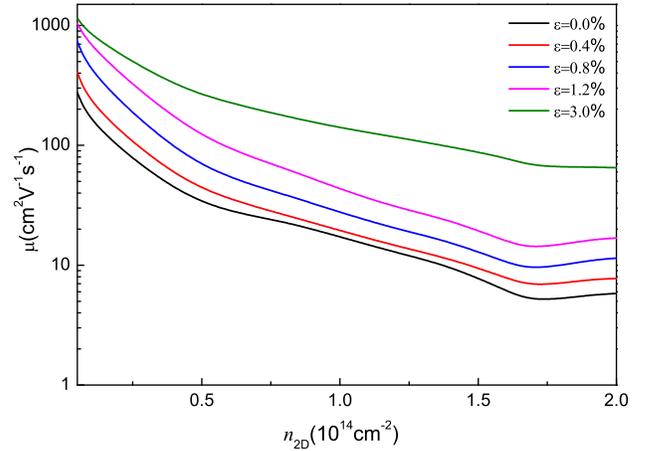}
\caption{(Color online) The mobility ${\mu}$ as a function of the doping concentration $n_{\rm{2D}}$ at various strains of $\varepsilon$ = 0\%, 0.4\%, 0.8\%, 1.2\%, and 3.0\%. \label{fig:4}}
\end{figure}

Figure~\ref{fig:3}(a) shows the transport spectral function $\alpha^2_{tr}\rm{F}(\omega)$ versus doping concentration under different strains. $\alpha^2_{tr}\rm{F}(\omega)$ spreads through a wide frequency range, but shows strong peaks at low frequencies, 80 $\sim$ 150 $\rm{cm}^{-1} $, corresponding to the $\text{M}_{\rm p}$ phonons. The position of this main region shifts down with increase of the doping concentration, consistent with the behavior of the $\text{M}_{\rm p}$ phonons softening, while position of the high frequency region is almost unchanged. After the application of strain, $\alpha^2_{tr}\rm{F}(\omega)$ is then greatly suppressed. The curves of $\lambda_{tr}$ with electron doping concentration under different strains are plotted in   Fig.~\ref{fig:3}(b). One can see that in the absence of strain, $\lambda_{tr}$ is strongly dependent on the doping concentration, in agreement with the results of Raman spectroscopy and superconductivity researchs~\cite{Chakraborty2012,ye2012,ge2013}.  The maximum ($\sim$1.6) of $\lambda_{tr}$ appears at $n_{\rm{2D}} = 1.7 \times 10^{14}$ cm$^{-2}$. However, with added strain, the maximum of $\lambda_{tr}$ shifts down to 0.14, and the overall $\lambda_{tr}$ versus $n_\text{2D}$ curve becomes rather flat. The doping and strain effects on $\lambda_{tr}$ are consistent with the discussion of $\lambda^{\mathbf{q}}_{tr}$ as above illustration.

\subsection{Mobility}

The mobility ${\mu}(T)$ calculated at room temperature ($T = 300$ K) are plotted in Fig.~\ref{fig:4}. Under the strain-free condition, $\mu$ is inversely proportional to the doping concentration in a wide range of doping level and increases slightly in the high doping level due to the change of electron-phonon coupling. In the two previous theoretical studies~\cite{Kaasbjerg2012,li2013}, the mobility has been estimated to be 130 (410) cm$^2$/(V$\cdot$s) respectively, depending on with (without) the consideration of the scattering of Q valley. It agrees well that our estimation of 275 cm$^2$/(V$\cdot$s) at $n_{\rm{2D}}=5 \times 10^{12}$ cm$^{-2}$ falls somewhere between them.

Next, we turn to discuss the stain effect on the mobility. At the low doping concentration of $n_{\rm{2D}}=5\times{10}^{12}\rm{cm}^{-2}$, $\varepsilon$ = 3.0\% strain can increase mobility from 275 to 1150 cm$^2$/(V$\cdot$s), for the strain can greatly suppress the electron-phonon coupling. In addition, the change of mobility by strain from $\varepsilon$ = 1.2\% to $\varepsilon$ = 3.0\% is inconspicuous in the low doping level ( $ < {10}^{13}\rm{cm}^{-2}$). It can be ascribed to that the influence of Q valley scattering to the mobility is rather weak such that can be almost totally removed by a small strain ($ > 1.2\%$). At the high doping concentration ($1.7\times{10}^{14}\rm{cm}^{-2}$), the modulation of mobility by strain becomes more significant, from 4.9 to 66.0 cm$^2$/(V$\cdot$s) by a factor of 10 [Fig.~\ref{fig:4}].

Although the present estimation of 275 cm$^2$/(V$\cdot$s) is coincide with theoretical predictions and some experiments as well, a significant difference was observed compared with the high value of 700 cm$^2$/(V$\cdot$s)~\cite{das2013}. Considering the lattice mismatching with the substrates in the different experimental preparations, the strain appears inevitably in $\rm{MoS}_{2}$. Nevertheless, it is provided in our results that the mobility is exquisitely sensitive to the external strain, $\sim$ 1000 cm$^2$/(V$\cdot$s) improved by the strain of 1.2\%. Clearly, the question about disparity between theoretical predictions and experimental results can be reasonably illustrated in this work.

\section{Summary}
In summary, we have studied the doping and strain effects on the transport electron-phonon coupling and intrinsic mobility in monolayer $\rm{MoS}_{2}$ based on first-principles methods with Boltzmann transport theory. We estimate that under the strain-free condition the mobility has a maximum 275 cm$^2$/(V$\cdot$s). Based on the analysis of electron-phonon coupling, it is found that the inter-valley scattering between the Q and K valleys assisted by $\rm{M_p}$ phonons are the main source of electric resistance. However, applying a small amount of strain can significantly increase the mobility by several times for the modulation effects on the lowest conduction band. Our results illustrate the effective modulation of strain on mobility in $\rm{MoS}_2$.

\begin{acknowledgments}
This work was supported by the MOST Project of China (Grants Nos. 2014CB920903 and 2011CBA00100), the NSFC (Grant Nos.11225418, 11174337 and 11374033), the Specialized Research Fund for the Doctoral Program of Higher Education of China (Grant No. 20121101110046 and 20131101120052), the Excellent Young Scholars Research Fund (Grant No. 2013CX04004) and the Basic Research Fund (Grant No. 20121842009) of Beijing Institute of Technology. DX is supported by the U.S. Department of Energy, Office of Basic Energy Sciences, Materials Sciences and Engineering Division.
\end{acknowledgments}

\setcounter{figure}{0}
\renewcommand{\thefigure}{A\arabic{figure}}
\appendix
\section{Nesting Function}
\begin{figure}[H]
\includegraphics[width=0.5\textwidth,height=3in]{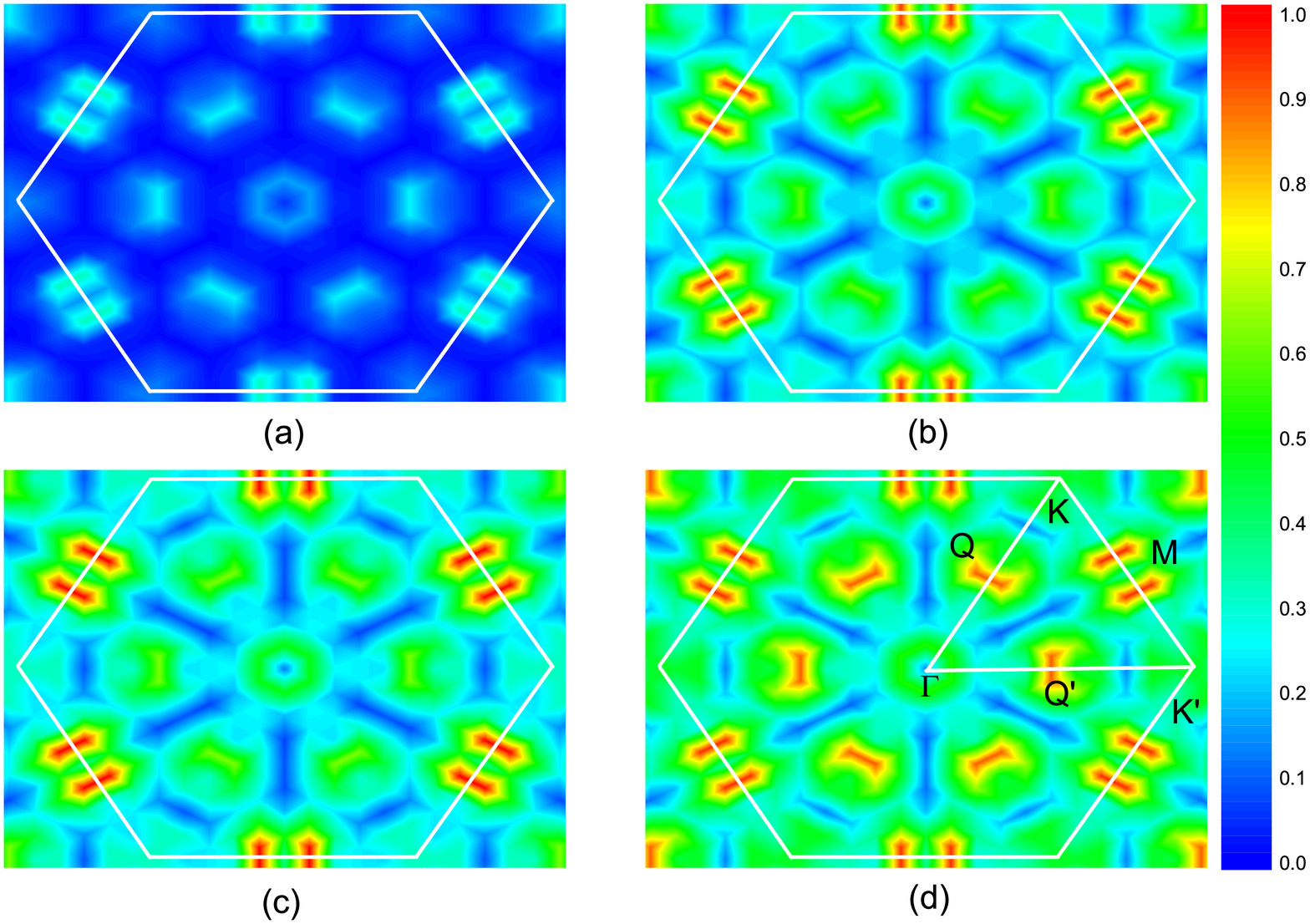}
\caption{(Color online) The map distributions of nesting function $X_{\mathbf{q}}$ at different doping concentrations $n_{\rm{2D}} = 0.6(\rm{a})$, $1.6(\rm{b})$, $1.7 (\rm{c})$ \rm{and } $1.9(\rm{d}) \times 10^{14}$ cm$^{-2}$ with strain-free condition, respectively.
\label{fig:5}}
\end{figure}

In this Appendix, we illustrate two additional channels of electronic transitions, $\Gamma_{\rm e}+ \text{Q}_{\rm p}(\text{K}_{\rm p}) \longleftrightarrow\text{Q}_{\rm e} (\text{K}_{\rm e})$, resulting from the lowering of conduction band at the $\Gamma$ point when the doping concentration increases from 1.7 to 1.9 $\times 10^{14}$ cm$^{-2}$. For clear observation, we calculated the nesting function $X_\mathbf q$,

\begin{eqnarray}
X_{\mathbf{q}}&=&\frac{1}{N_k}{\sum_{\substack{\mathbf{k},m,n}}}\delta(\epsilon_{\mathbf{k},m}-\epsilon_{F})
\delta(\epsilon_{\mathbf{k}+\mathbf{q},n}-\epsilon_{F}),
\label{eq:xq}
\end{eqnarray}
which describes the geometrical property of the Fermi surface. Figure~\ref{fig:5} shows the change of $X_{\mathbf{q}}$ at different doping concentrations. In the doping level of $n_{\rm{2D}}<$ 1.7 $\times 10^{14}$ cm$^{-2}$, $X_\mathbf q$ around the Q (K) point has no obvious change [Figs.~\ref{fig:5}(a-c)]. But one can clearly see that $X_\mathbf q$ around the Q (K) point enlarges significantly when $n_{\rm{2D}}$ increases to 1.9 $\times 10^{14}$ cm$^{-2}$ as shown in Fig.~\ref{fig:5}(d). Combining with the evolution of band structure along with increasing $n_{\rm{2D}}$ from 1.7 to 1.9 $\times 10^{14}$ cm$^{-2}$ [Fig.~\ref{fig:1}(c)], one can deduce that there appears two additional channels of electronic transitions, i.e., $\Gamma_{\rm e}+\text{Q}_{\rm p} (\text{K}_{\rm p}) \longleftrightarrow\text{Q}_{\rm e} (\text{K}_{\rm e})$. However, due to the weak electron-phonon coupling of these two additional channels, they cannot be clearly seen in Fig.~\ref{fig:2}(d).  Therefore, the dominated channel of electronic transition is $\text{K}_{\rm e}+\text{M}_{\rm p}\longleftrightarrow\text{Q}_{\rm e}$, as discussed in the main text.

\end{document}